\documentclass[12pt]{article}

\input epsf
\usepackage{amssymb,amsmath,wrapfig,subfigure}
\usepackage[matrix,arrow,curve]{xy}
\usepackage{cite}
\usepackage{graphicx,color}
\usepackage[debug,pageanchor=false]{hyperref}
\definecolor{link}{rgb}{.8,.15,.1}
\hypersetup{colorlinks=true,linkcolor=link,citecolor=link,urlcolor=link,linktocpage}

\makeatletter
\@addtoreset{equation}{section}
\makeatother

\def\rr {{\Bbb R}}
\def\cc {{\Bbb C}}

\def\del {\partial}

\def\del {\partial}

\def\sla#1{\rlap{\begin{picture}(10,10)
\put(0,0){\line(1,1){10}}
\end{picture} }#1}

\begin{document}

       \begin{titlepage}

       \begin{center}

       \vskip .3in \noindent

       {\Large \bf{Supersymmetry on Curved Spaces and Holography}}

       \bigskip

	 Claudius Klare, Alessandro Tomasiello and Alberto Zaffaroni\\

       \bigskip
		 Dipartimento di Fisica, Universit\`a di Milano--Bicocca, I-20126 Milano, Italy\\
       and\\
       INFN, sezione di Milano--Bicocca,
       I-20126 Milano, Italy

       \vskip .5in
       {\bf Abstract }
       \vskip .1in

       \end{center}

We study superconformal and supersymmetric theories on Euclidean four- and three-manifolds with a view toward holographic applications.
Preserved supersymmetry for asymptotically locally AdS solutions implies the existence of a (charged) ``conformal Killing spinor'' on the boundary. 
We study the geometry behind the existence of such spinors. We show in particular that, in dimension four, they exist on any complex manifold. This implies that a superconformal theory has at least one supercharge on any such space, if we allow for a background field (in general complex) for the R-symmetry. We also show that this is actually true for any supersymmetric theory with an R-symmetry. We also analyze the three-dimensional case and provide examples of supersymmetric theories on Sasaki spaces.

       \noindent

       \vfill
       \eject


       \end{titlepage}

\section{Introduction} 
\label{sec:intro}

When studying quantum field theory on curved spacetime, symmetries play an important role. For example, in absence of a Killing vector, there is no canonical choice of vacuum for the theory. For supersymmetric theories, a natural question is whether any of the fermionic symmetries that the theory enjoys on flat spacetime still hold on a given curved manifold $M$. When this is the case, one can still consider some of the usual techniques from supersymmetry to study the theory; the curvature of $M$ acts then as an infrared regulator. 

It is natural to expect that spaces which have many bosonic symmetries should also preserve many supercharges. For example, there has recently been much progress in understanding (Euclidean) supersymmetric field theories using localization techniques, on $S^4$ \cite{pestun-S4}, $S^3$ \cite{kapustin-willett-yaakov,jafferis-Z,hama-hosomichi-lee,nawata}, and various squashed $S^3$ \cite{hama-hosomichi-lee,hama-hosomichi-lee2,imamura-yokoyama}. Moreover, a general way to write a supersymmetric field theory on a curved space has recently been proposed in \cite{festuccia-seiberg}. It consists in coupling the field theory to supergravity, and in subsequently sending $M_{\rm Pl}\to \infty$ while freezing the supergravity fields to some background value. In many interesting Euclidean cases, the background values of the auxiliary fields are complex.  

Superconformal theories can also be studied using holography. Most of the applications of holography so far have been devoted to conformally flat manifolds such as Minkowski space, $S^{d-1} \times S^1$, and $S^d$ in the Euclidean case. In light of the recent progress mentioned above, it is interesting to study in more detail how holography works for more general manifolds.\footnote{For example, \cite{martelli-passias-sparks,martelli-sparks-nuts} have recently considered ${\cal N}=2$ theories on squashed $S^3$, reproducing some of the localization results in \cite{hama-hosomichi-lee2,imamura-yokoyama}. For older examples in Lorentzian signature, see \cite{gauntlett-gutowski-suryanarayana,behrndt-klemm}.} Only some of the bosonic and fermionic symmetries of the theory will be present, but one expects the symmetries of the theory to be matched by isometries of the gravity dual. In particular, the supersymmetry transformations in the bulk induce the transformations of conformal supergravity when restricted to the boundary \cite{balasubramanian-gimon-minic-rahmfeld}.

In this paper, we focus on the case with one preserved supercharge in Euclidean signature. Starting from very mild assumptions on the structure of the gravity dual, we point out that the boundary manifold will have a \emph{conformal Killing spinor}, namely a solution of the conformally invariant equation $\left(\nabla^A_m - \frac1d \gamma_m D^A\right)\epsilon=0$, where $D=\gamma^m \nabla_m$ is the Dirac operator, and ${}^A$ denotes twisting by a gauge field, in general complex. Following the same logic as in \cite{festuccia-seiberg}, applied this time to conformal supergravity \cite{kaku-townsend-vannieuwenhuizen1,kaku-townsend-vannieuwenhuizen2,kaku-townsend-vannieuwenhuizen3,ferrara-zumino-conformal}, the existence of a conformal Killing spinor is exactly the condition one needs in order to preserve one supercharge for a superconformal theory\footnote{The relevance of conformal Killing spinors was also realized for superconformal $\sigma$ models in \cite{sezgin-tanii,bergshoeff-cecotti-samtleben-sezgin}.}. 

In four dimensions, we show that a conformal Killing spinor exists on any complex manifold (K\"ahler or not)\footnote{There exist many complex manifolds which are not K\"ahler. The most famous example is perhaps the Hopf surface, which is diffeomorphic to $S^3\times S^1$, which will be considered in section \ref{sec:3d}, or more generally all ``class VII'' Kodaira surfaces. Primary and secondary Kodaira surfaces in the Enriques--Kodaira classification are also non-K\"ahler (see for example \cite[Ch.~VI]{barth-peters-vandeven}).}. Thus, a superconformal field theory on any complex manifold preserves at least one supersymmetry, if we turn on a background field (in general complex) for the R-symmetry. In fact, conformal supergravity gives rise to ordinary supergravity once one breaks conformal invariance by giving expectation value to fields in auxiliary compensator multiplets (as for example in \cite{kaku-townsend}; for a review see \cite{vanproeyen-conformal}).  In particular, applying this idea to a tensor multiplet gives rise to ``new minimal supergravity'' \cite{sohnius-west}. This suggests that one might extend our results to any supersymmetric theory with an R-symmetry (not necessarily conformal); and we indeed show, using again the method in \cite{festuccia-seiberg}, that a supersymmetric theory preserves at least one supercharge on any complex manifold. A very similar analysis has been presented for the ``old minimal supergravity'' in \cite{samtleben-tsimpis} \footnote{See also \cite{jia-sharpe}.}.  

Many of the interesting results we mentioned earlier about supersymmetric field theory on curved spaces are in three dimensions, and we consider this case too, after dimensionally reducing the equations for four-dimensional supergravity. We reproduce the known examples in \cite{hama-hosomichi-lee,hama-hosomichi-lee2,imamura-yokoyama}, but we also show that a theory preserves at least one supercharge on any Sasaki manifold. 

Our analysis can also be used to identify concretely the Lagrangian of the theory dual to a given supergravity background. Suppose we have a supergravity theory whose AdS solution is dual to a given CFT in flat space. If one has another solution of the same supergravity theory which is asymptotically locally AdS, it is possible to read off the value of the boundary metric and of the background field for the R-symmetry, and to write the Lagrangian of the CFT on the resulting curved space using our discussion in section \ref{sub:csugra} (and in particular (\ref{eq:cscoupling})), in agreement with the standard AdS/CFT dictionary. Moreover, from (\ref{eq:vdef}) one can also identify the background field appearing in new minimal supergravity, which is crucial to write the Lagrangian for any supersymmetric but non-conformal deformation of the CFT.

This paper is organized as follows. In section \ref{sec:ads}, we review how conformal Killing spinors arise from holography. In section \ref{sec:cks}, we study the geometry of conformal Killing spinors; in the charged case, we find that any complex manifold admits one. In section \ref{sec:newmin}, we show that any supersymmetric field theory preserves one supercharge on a complex manifold. Finally, in section \ref{sec:3d}, we consider the three-dimensional case.  

{\bf Note:} While completing this work we became aware of \cite{dumitrescu-festuccia-seiberg}, which has some overlap with our work. We are grateful to the authors of \cite{dumitrescu-festuccia-seiberg} for exchanging drafts with us before publication.

 
\section{Asymptotically locally AdS and superconformal theories} 
\label{sec:ads}

In this section, we review how supersymmetry in the bulk implies the existence of a ``conformal Killing spinor'' on the boundary  \cite{balasubramanian-gimon-minic-rahmfeld}; a very similar version of this computation has also appeared in \cite[App.~E]{cheng-skenderis}.
We will describe this for four-dimensional gravity in section \ref{sub:43}, and in section \ref{sub:54} for five-dimensional gravity. In section \ref{sub:csugra}, we will interpret the result in terms of the superconformal theory at the boundary. 

\subsection{From four-dimensional gravity to CFT$_3$'s} 
\label{sub:43}

Our starting point is an ${\cal N}=2$ gauged supergravity with an ${\rm AdS}_4$ vacuum corresponding to the dual  of a three-dimensional conformal field theory on flat space. According to the holographic dictionary, other solutions of the bulk theory
which are asymptotically AdS describe deformations (or different vacua) of the CFT.  We are interested in studying the CFT on a curved Riemannian manifold $M_3$ and therefore we look for solutions of the bulk theory with conformal boundary\footnote{Indices $M,N,\ldots$ are curved in the bulk; $m,n,\ldots$ will be curved indices on the boundary; $a,b,\ldots$ will be flat indices on the boundary.} $M_3$:
\begin{equation}
	ds^2_4 = \frac{dr^2}{r^2}+ ( r^2 ds^2_{M_3} + O(r))\ .
\end{equation}
In general, in order to define the theory on the curved manifold in a supersymmetric way, we will need to turn on a non trivial background for the  R symmetry current.
This corresponds to  a relevant deformation $A_m J^m$  of the CFT  and we expect  a non trivial profile of the graviphoton field in the bulk.
On the other hand, we do not want to include explicit deformations induced by scalar operators so we can safely assume that all the scalars in the bulk vanish at the boundary.
The supersymmetry variation of the gravitino then reduces near the boundary to the form
\begin{equation}\label{eq:nabepsbulk}
	\left(\nabla^A_M  + \frac12 \gamma_M + \frac i 2 \sla{ F} \gamma_M\right) \epsilon = 0 
\end{equation}
where 
\begin{equation}
	\sla F \equiv \frac12 F_{MN} \gamma^{MN}\ ,
\end{equation}
and
\begin{equation}\label{eq:nablaA}
	\nabla^A_M \equiv \nabla_M - i A_M \ .
\end{equation}
Our boundary condition requires $A_M$ to have only components along $M_3$ and to be independent of $r$ near the boundary, which is compatible with the equations of motion. From the point of view of the gravity solution, this corresponds to the non-normalizable mode for $A$, which indeed is interpreted in AdS/CFT as the deformation of the theory induced by a background field for the R-symmetry. We can also turn on the bulk fields corresponding to global symmetries of the CFT but, allowing for a redefinition in $A$,
this will not change the form of the supersymmetry transformation.
 
The behavior of  $\sla F$ is of order $O(r^{-2})$. So at leading order we can neglect its contribution to (\ref{eq:nabepsbulk}). In frame indices $(a,4)$, $a=1,2,3$, we get
\begin{equation}\label{eq:4dbulk}
	\left(\del_4 + \frac 12 \gamma_4\right) \epsilon=0 \ ,\qquad
	\left(\nabla^A_a + \frac r 2 \gamma_a (1+\gamma_4) \right) \epsilon= 0 \ , 
\end{equation}
where $\nabla_a^A$ is now the covariant derivative with respect to the metric $ds^2_{M_3}$. In the second equation in (\ref{eq:4dbulk}), a term from the covariant derivative relative to the metric $ds^2_4$ has combined with the term $\frac12 \gamma_a$ in (\ref{eq:nabepsbulk}).

Since $\gamma_4$ squares to one, we can divide spinors into eigenspaces of eigenvalue $\pm 1$, $\gamma_4 \epsilon_\pm = \pm \epsilon_\pm$. The first equation in (\ref{eq:4dbulk}) then gives 
\begin{equation}
	\epsilon= r^{\frac12} \epsilon_- + r^{-\frac12} \epsilon_+\ .
\end{equation}
Plugging this into the second equation of (\ref{eq:4dbulk}) gives, at leading order, 
\begin{equation}
	\nabla^A_a \epsilon_- + \gamma_a \epsilon_+ = 0 \ .
\end{equation}
We can use $\gamma^4$ to reduce spinors from four dimensions to three. In a basis where
\begin{equation}\label{eq:4dgamma}
	\gamma^a = 
	\left(
	\begin{array}{cc}
		0 & \sigma^a \\ \sigma^a & 0 
	\end{array}
	\right)
	\ ,\qquad
	\gamma^4=
	\left(
	\begin{array}{cc}
		0 & i \\ -i & 0 
	\end{array}
	\right) \ ,
\end{equation} 
the spinors $\epsilon_\pm$ can be rewritten as 	$\epsilon_\pm =
\bigl( \begin{smallmatrix}
			\pm i \chi_\pm \\ \chi_\pm  
\end{smallmatrix}\bigr)$, where $\chi_\pm$ are three-dimensional spinors. This gives 
\begin{equation}
	\nabla^A_a \chi_- = -i \sigma_a \chi_+ \ . 
\end{equation}
We can actually derive $\chi_+$ by taking the trace: 
\begin{equation}
	\chi_+ = \frac i3 D^A  \chi_- \ ,  
\end{equation} 
where $D^A \equiv \sigma^a\nabla_a $ is the Dirac operator. We have obtained
\begin{equation}\label{eq:cks3}
	\left(\nabla^A_a - \frac13 \sigma_a D^A \right) \chi_- = 0 \ .
\end{equation}
A solution to this equation is known as a (charged) \emph{conformal Killing spinor}, or as a \emph{twistor spinor}. We will review the mathematics behind it and classify its solutions in section \ref{sec:cks}. 


\subsection{From five-dimensional gravity to CFT$_4$'s} 
\label{sub:54}

The analysis of ${\cal N}=2$ gauged supergravity in five dimensions is similar and we will be brief. For the same reason as in four dimensions we only keep the graviphoton and discard terms with the curvatures
in the supersymmetry variations which reduce to
\begin{equation}\label{eq:5dvar}
\nabla_M \epsilon^I  + \left ( \frac i 2 \gamma_M -  A_M\right) \epsilon^{IJ} \epsilon^J = 0\, \qquad I,J=1,2  \ .
\end{equation}
In the Lorentzian-signature theory, $\epsilon^I$ are symplectic-Majorana spinors. In the Euclidean case, we relax this condition. We use the gamma matrices defined in (\ref{eq:4dgamma}), with the addition of $\gamma_5= \gamma_{1234}$.

The component along $e^5= dr/r$ of equation (\ref{eq:5dvar}) gives
\begin{subequations}\label{eq:5dbulkr}
	\begin{align}
		\del_5 \epsilon^1 + \frac i 2 \gamma_5  \epsilon^2 &=  0 \ , \\
		\del_5 \epsilon^2 - \frac i 2 \gamma_5  \epsilon^1 &=  0 \ . 
	\end{align}
\end{subequations}
Combining the two equations we have $( \del_5^2 - 1/4 )\epsilon^1=0$ and therefore
\begin{equation}\label{eq:5dbehav}
 	\epsilon^1= r^{\frac12} \epsilon + r^{-\frac12} \eta\ .
\end{equation}
Plugging  this expression into the other components of (\ref{eq:5dvar}) and eliminating   $\epsilon^2$ 
using the second equation of (\ref{eq:5dbulkr}) we obtain, at leading order, 
\begin{equation}\label{eq:cks4}
	\left ( \nabla_a  - i A_a \gamma_5\right )\epsilon  + \gamma_a \gamma_5 \eta = 0 \ ,
\end{equation}
where now $a=1,\ldots,4$, and the covariant derivative and all other quantities are taken with respect to the four-dimensional manifold $M$; $\epsilon$ and $\eta$ 
are four-dimensional Dirac spinors.  

We can also separate the previous equation according to four-dimensional chirality ($\gamma_5=\pm 1$):
\begin{equation}
	\begin{split}
		\nabla_a^A  \epsilon_+   &=    \gamma_a  \eta_- \ , \\
	 \nabla_a^A  \epsilon_-  &=   - \gamma_a  \eta_+  \ ;
	\end{split}
\end{equation}
and, by eliminating $\eta_\pm$, we find the equation for conformal Killing spinors
\begin{equation}\label{eq:cks42}
 \nabla_a^A  \epsilon_\pm   = \frac14   \gamma_a  D^A \epsilon_\pm\ ,
\end{equation} 
where $\nabla_a^A\epsilon_\pm = (\nabla_a \mp i A_a)\epsilon_\pm$.

As usual, in the Euclidean the spinors have been doubled. A computation in the Minkowskian case would give a similar result but with
$\epsilon$ and $i \eta$ Majorana spinors. 


\subsection{Conformal Killing spinors and superconformal theories} 
\label{sub:csugra}

The appearance of the equation (\ref{eq:cks3}) and (\ref{eq:cks42}) for a conformal Killing spinor at the conformal boundary of the gravity solution can be easily explained. 

Recall first that, if we want to define a supersymmetric theory on a curved manifold $M$, an efficient strategy \cite{festuccia-seiberg} consists in coupling the theory to supergravity and then freeze the fields of the gravitational multiplet. The value of the auxiliary fields determines the coupling of the theory to the curved background. 

For a superconformal theory, one can proceed similarly and couple the theory to the fields of conformal supergravity $g_{mn}$, $\psi_m$ and $A_m$.  At the linearized level, these fields couple to the superconformal currents:
\begin{equation}\label{eq:cscoupling}
	-\frac 12 g_{mn}T^{mn} + A_m J^m + \bar \psi_m {\cal J}^m \ ,
\end{equation}
where $J^m$ is the R-symmetry and ${\cal J}^m$ is the supersymmetry current. 
For us, the fields of conformal gravity will play the role of background fields; since we work in Euclidean signature, we will allow the auxiliary field $A_m$ to be complex. 

In order to preserve some supersymmetry, the gravitino variation must vanish. For simplicity, we write the variation
for a four-dimensional theory where they read (with obvious redefinitions) \cite{kaku-townsend-vannieuwenhuizen3,vanproeyen-conformal}
\begin{equation}\label{eq:supconf}
\delta \psi_m =\left ( \nabla_m  - i A_m \gamma_5\right )\epsilon  + \gamma_m \gamma_5 \eta 
\end{equation}
where $\epsilon$ is the parameter for the  supersymmetries $Q$ and $\eta$ for the superconformal transformations $S$. It is crucial for our arguments that, as stressed many times in the old days\footnote{See for instance
\cite{das-kaku-townsend} for a very simple example of the logic we will be using in this paper.}, the algebra of the superconformal transformations of $g_{mn},\psi_m, A_m$ closes off shell.  Therefore  the 
variation (\ref{eq:supconf})  depends only  on the background field $A_m$ and is not modified by the coupling to matter. Moreover, the supergravity action for the fields $g_{mn},\psi_m, A_m$ is separately
invariant and can be safely omitted without spoiling the superconformal invariance of the matter part. 

The vanishing of the gravitino variation constrains the manifolds where we can have supersymmetry.
As expected, equation (\ref{eq:supconf}) is identical  to  (\ref{eq:cks4}) which, in turn,  is equivalent to the conformal Killing equation. Notice  that $\epsilon$ and $\eta$ in our bulk computation appear in the asymptotic 
expansion (\ref{eq:5dbehav}) with a different power of $r$ corresponding precisely to the conformal dimension of  the supercharges $Q$ and $S$.



\section{Geometry of conformal Killing spinors} 
\label{sec:cks}

In this section, we will review some geometry behind the conformal Killing spinor equation
\begin{equation}
	P^A_m \epsilon \equiv \left (\nabla_m ^A - \frac1d \gamma_m D^A\right ) \epsilon =0 \, , 
\end{equation}
and classify its solutions.  Notice  that the conformal Killing operator $P^A_m$   is covariant under Weyl rescaling. The operator $\bar P^A_m$ of the rescaled metric $\bar g = e^f g$ is indeed given by
\begin{equation}
\bar P^A_m e^{f/4} = e^{f/4} P^A_m\ .
\end{equation}

\subsection{The $A=0$ case} 
\label{sub:cksa0}

In the uncharged case ($A=0$), the conformal Killing spinor equation reads
\begin{equation}\label{eq:cks}
	\left(\nabla_m - \frac1d \gamma_m D\right) \epsilon \equiv P_m \epsilon = 0 \ .
\end{equation}
One way to think of the operator $P_m$ is the following. The covariant operator $\nabla_m$ goes from the bundle of spinors $\Sigma$ to the bundle $T \otimes \Sigma$ of vector-spinors. The sections of the latter  are a reducible representation of the orthogonal (or Lorentz) group. It can be written as the direct sum of two representations: a ``trace'', defined by taking a section $\psi_m$ and multiplying it by $\gamma^m$, and the traceless part.\footnote{This is familiar from the NS$\otimes$R sector of the NSR superstring, which decomposes into a dilatino (the trace) and the gravitino (the traceless part).} The orthogonal projector  on this second irreducible representation can be written as $\delta_m^n - \frac1d \gamma_m \gamma^n$. Now, projecting $\nabla_m$ on the trace representation gives the Dirac operator $D$, while projecting on the traceless part gives 
\begin{equation}
	\left(\delta_m^n - \frac1d \gamma_m \gamma^n\right) \nabla_n = \nabla_m - \frac1d \gamma_m D = P_m \ .
\end{equation}

So in a sense $P_m$ is the ``complement'' of the Dirac operator. Some of the properties of $P_m$ (and of its zero modes, the conformal Killing spinors) have been studied by mathematicians; see for example \cite{lichnerowicz} in the Euclidean case and \cite{baum-CKS} in the Lorentzian case. In particular, some of these results can be used to classify completely the manifolds on which a conformal Killing spinor can exist, as we will now review.

Consider a conformal Killing spinor $\epsilon$. One can show that $D^2 \epsilon \propto R\,\epsilon$, where $R$ is the scalar curvature. Using the solution to the Yamabe problem \cite{lee-parker,schoen}, one can make $R$ constant by a conformal rescaling of the metric. $\epsilon$ is then an eigenspinor for $D^2$; namely,
\begin{equation}
	(D^2- \mu^2) \epsilon = (D-\mu)(D + \mu) \epsilon = 0 \ ,
\end{equation}
for some $\mu$. The spinors
\begin{equation}
	\psi \equiv \epsilon + \frac1{\mu} D \epsilon \ ,\qquad 
	\tilde \psi \equiv \epsilon - \frac1{\mu} D \epsilon \ ,
\end{equation}
are then eigenspinors of $D$. A theorem by Hijazi \cite{hijazi} now tells us that any eigenspinor of $D$ is also a \emph{Killing spinor}, namely a spinor $\epsilon$ that satisfies
\begin{equation}\label{eq:ks}
	\nabla_m \epsilon = \mu \gamma_m \epsilon \ .
\end{equation}
Such spinors are familiar from the supergravity literature; for example, one can find explicit expression for Killing spinors on the sphere $S^n$ in \cite{lu-pope-rahmfeld}. One can readily check that every Killing spinor is a conformal Killing spinor; thus, a priori (\ref{eq:ks}) would seem to be more restrictive than (\ref{eq:cks}). However, as we have just described, existence of a solution to (\ref{eq:cks}) is in fact equivalent to existence of a solution to (\ref{eq:ks}) (with a Weyl rescaled metric). 

In fact, manifolds which admit Killing spinors have been classified. Notice first that the usual compatibility between different components of (\ref{eq:ks}) gives $R_{m n}= -2 \mu^2 g_{mn}$. This implies that $\mu$ should be either real or purely imaginary. The real case can be shown \cite{baum} to be realized only on non-compact manifolds, which are in fact a warped product of $\rr$ with any manifold $M$, with metric $dr^2 + e^{-4 \mu r} ds^2_M$. When $\mu$ is purely imaginary, one can observe \cite{baer} that the existence of a Killing spinor on $M_d$ implies the existence of a covariantly constant spinor on the cone $C(M_d)$. Such manifolds are in turn classified using their restricted holonomy. 

For example, in dimension four, the cone $C(M_4)$ would be a five-dimensional manifold with restricted holonomy, which can only be $\rr^5$. This tells us that $S^4$ is the only four-manifold with Killing spinors, and thus the only four-manifold with conformal Killing spinors (up to Weyl rescaling). 
  
The case of $S^4$ is also instructive in other respects. It is known that there is no almost complex structure on this manifold. A chiral spinor defines at each point an almost complex structure; thus, there can be no chiral spinor without zeros on $S^4$. On the other hand, a Killing spinor has no zeros, because (\ref{eq:ks}) implies that the norm of $\epsilon$ is constant. There is no contradiction: a Killing spinor is never chiral; so $\epsilon= \epsilon_+ + \epsilon_-$, where $\epsilon_\pm$ are chiral. Both $\epsilon_+$ and $\epsilon_-$ have one zero, which explains why there is no almost complex structure on $S^4$, but the norm of $\epsilon$ is still constant. In fact, \cite[Th.~7]{lichnerowicz} shows that, in any  dimension, the sphere $S^d$ is the only manifold on which a conformal Killing spinor can have a zero. 

In dimension three the situation is similar. Since  $\rr^4$ is the only four-manifold with restricted holonomy, $S^3$ (or quotients thereof) is the only compact three-manifold with Killing spinors. Unlike in  the four dimensional case, the conformal Killing spinors on $S^3$ never vanish.  In higher dimensions, we have a larger class of possibilities.   The existence of Killing spinors identifies Sasaki-Einstein manifolds in five dimensions and nearly-K\"ahler manifolds in six \cite{baer}. The corresponding cones with restricted holonomy are Calabi-Yau three-folds and $G_2$ manifolds, respectively. Only in the case of $S^6$ the conformal Killing  spinor is allowed to have a zero. 
 
Let us summarize the results we have reviewed in this subsection. If a manifold admits a conformal Killing spinor (uncharged, namely with $A=0$), it also admits a closely related Killing spinor. Manifolds with Killing spinors, in turn, are also completely classified; they are either warped products of a manifold with $\rr$, or bases of cones with covariantly constant spinors.


\subsection{The $A\neq 0$ case in four dimensions} 
\label{sub:cksa}

We will now turn to the case with $A\neq 0$:
\begin{equation}\label{eq:cksa}
	\left(\nabla^A_m - \frac1d \gamma_m D^A\right) \epsilon = P^A_m \epsilon = 0 \ ,\qquad m=1,\ldots,4\ .
\end{equation}
In general, recalling (\ref{eq:nablaA}), $A$ will be a connection on a bundle. We will actually take ${\rm Re} A$ to be a connection on a U(1) bundle ${\cal U}$, and ${\rm Im} A$ to be a one-form. Accordingly, $\epsilon$ will be not quite a spinor, but a ``charged'', or Spin$^c$, spinor; namely, a section of
\begin{equation}
	{\cal U}\otimes \Sigma\ ,
\end{equation}
where $\Sigma$ is the spinor bundle. 

(\ref{eq:cksa}) has also been considered by mathematicians (see e.g.~\cite[Part III]{lichnerowicz2}), but in this case it is no longer true that existence of its solutions is equivalent to the existence of charged Killing spinor (which have been studied for example in \cite{moroianu}). Thus a complete classification of the solutions to (\ref{eq:cksa}) is currently not available.

We will thus study (\ref{eq:cksa}) here. In this section, we will deal with the four-dimensional case. Since (\ref{eq:cksa}) does not mix different chiralities (unlike (\ref{eq:ks})), we can consider its chiral solutions separately. For simplicity, we will assume that $\epsilon=\epsilon_+$ is a spinor of positive chirality. 

\subsubsection{Intrinsic torsions} 
\label{ssub:intr}

We can borrow some of the tools that have been successfully used in the analysis of supersymmetric solutions in supergravity. The first idea is to parameterize the covariant derivatives of $\epsilon_+$ in terms of a basis of spinors. This strategy has been used for a long time (for example \cite[(2.2)]{gmpt}); in the case of a four-dimensional Euclidean manifold, this was used recently in \cite{lust-patalong-tsimpis,samtleben-tsimpis}. In the case at hand, a basis in the space of spinors with positive chirality is given by 
\begin{equation}\label{eq:+basis}
	\epsilon_+ \ ,\qquad \epsilon_+^C \equiv C \epsilon^*\ ,
\end{equation} 
where $C$ is the intertwiner such that $\gamma_m^*= C^{-1} \gamma_m C$. A basis for the space of spinors of negative chirality is given by either 
\begin{equation}\label{eq:-basis}
	\gamma_m \epsilon_+
\end{equation}
or by
\begin{equation}\label{eq:-basisC}
	\gamma_m \epsilon_+^C\ ;
\end{equation}
as we will see shortly, this second choice is related to (\ref{eq:-basis}). Using the basis (\ref{eq:+basis}), we can expand
\begin{equation}\label{eq:pq}
	\nabla_m \epsilon_+ = p_m \epsilon_+ + q_m \epsilon^C_+\ .
\end{equation}
$p_m$, $q_m$ are locally complex one-forms. Globally speaking, ${\rm Im} p$ is a connection on ${\cal U}$, ${\rm Re} p$ is a one-form, and $q$ is a section of ${\cal U}^2\otimes T^*$. 

An alternative, perhaps more transparently geometrical, point of view, consists in noticing that $\epsilon_+$ defines an U(2) structure on $M_4$. We can express it in terms of forms by considering the bispinors 
\begin{equation}\label{eq:bisp}
	\epsilon_+ \otimes\epsilon_+^\dagger = \frac14 e^B e^{-i\,j}
	\ ,\qquad
	\epsilon_+ \otimes \overline{\epsilon_+} = \frac14 e^B \omega
	\ ,
\end{equation}
where $\overline{\epsilon}\equiv \epsilon^T C^{-1}$, $e^B\equiv || \epsilon_+ ||^2$, and $j$ is a real two-form. $\omega$ is locally a complex two-form; globally, it is actually a section of 
\begin{equation}
	{\cal U}^2\otimes \Omega^{2,0}\ ,
\end{equation}
where recall ${\cal U}$ is the U(1) bundle for which $A$ is a connection. $j$ and $\omega$ satisfy 
\begin{equation}\label{eq:SU2}
    \omega^2=0 \ ,\qquad \omega \wedge \bar \omega = 2 j^2 \ ,
\end{equation}
or, more symmetrically, 
\begin{equation}\label{eq:SU2q}
\begin{split}
	&j \wedge {\rm Re} \omega = {\rm Re} \omega \wedge {\rm Im} \omega = {\rm Im} \omega \wedge j = 0\ ,  \\
	&j^2=({\rm Re} \omega)^2 = ({\rm Im} \omega)^2\ . 
\end{split}
\end{equation}
One can use these forms to relate the two choices for a basis of spinors of negative chirality, (\ref{eq:-basis}) and (\ref{eq:-basisC}):
\begin{equation}\label{eq:geeC}
	\gamma_m \epsilon_+ = \frac12 \omega_{mn} \gamma^n \epsilon_+^C\ .
\end{equation}
Notice that this also implies that $\epsilon_+$ and $\epsilon_+^C$ are annihilated by half of the gamma matrices\footnote{In other words, they are \emph{pure}; it is indeed well-known that any spinor in even dimension $\le 6$ has this property \cite[Rem.~9.12]{lawson-michelsohn}.}: 
\begin{equation}\label{eq:proj}
	\bar\Pi_m{}^n \gamma_n\epsilon_+ = 0 
	\ ,\qquad
	\Pi_m{}^n \gamma_n \epsilon_+^C = 0 \ ,
\end{equation}
where $\Pi_m{}^n \equiv \frac12 ( \delta_m^n - i I_m{}^n) = \frac14 \omega_{mp}\bar \omega^{np}$ is the holomorphic projector (relative to the almost complex structure $I_m{}^n \equiv j_{mp}g^{pn}$).

Strictly speaking, the previous discussion should be taken with a grain of salt in the case where $\epsilon_+$ has  zeros. In general $j$ and $\omega$ will not be well-defined on any zero $z_i$, and will define a U(2) structure only on $M_4 - \{ z_i \}$.

Similarly to (\ref{eq:pq}), it is easy to parameterize the derivatives of $j$ and $\omega$, by decomposing $dj$ and $d \omega$ in SU(2) representations. A three-form $\alpha_3$ can always be written as $\alpha_3 = \alpha_1 \wedge j$, where $\alpha_1$ is a one-form; or, it can be decomposed into its $(2,1)$ part and its $(1,2)$ part, which can in turn be reexpressed as $\omega\wedge \beta_{0,1}$ and $\bar \omega \wedge \tilde\beta_{1,0}$. These two possibilities can be exchanged with one another by using (\ref{eq:geeC}).

We follow both strategies to write
\begin{equation}\label{eq:wi}
	d j = w^4 \wedge j \ ,\qquad
	d \omega = w^5 \wedge \omega + w^3 \wedge \bar\omega\ .
\end{equation}
$w^4$ is a real one-form. $w^5$ and $w^3$ are locally complex one-forms; globally, ${\rm Im} w^5$ is a connection on ${\cal U}^2$, ${\rm Re} w^5$ is a one-form, and $w^3$ is a section of ${\cal U}^4\otimes T^*$. 

The $w^i$ are collectively called ``intrinsic torsion''. Our choice to write $dj$ using $j$ and $d \omega$ using $\omega$ and $\bar \omega$, and our names for the one-forms $w^i$, might seem mysterious. We made these choices to be as close as possible to a notation commonly used for U(3) structures on six-manifolds, where the intrinsic torsion consists of forms $W_1,\ldots, W_5$ of various degrees (a notation which is also not particularly suggestive, but which has become traditional; see \cite{chiossi-salamon,gmpt}). Notice that $w_5$ can be assumed to have $(0,1)$ part only, and $w_3$ to have only $(1,0)$ part. 

Our parameterization of $\nabla \epsilon_+$ in (\ref{eq:pq}) is nothing but a spinorial counterpart of the intrinsic torsions $w_i$ in (\ref{eq:wi}). In fact, we can easily compute a relation between the two, using the definitions (\ref{eq:bisp}) of $j$ and $\omega$. We get: 
\begin{equation}
	w^4 = -2 {\rm Re} (\bar q \llcorner \omega) \ ,\qquad
	w^5_{0,1} = 2i ({\rm Im} p)_{0,1} -\frac12 q \llcorner \bar \omega\ ,\qquad
	w^3_{1,0} = \frac12 q \llcorner \omega \ .
\end{equation}
As a byproduct, we also obtain a relation on $B$ (which was defined earlier as $e^B\equiv || \epsilon_+ ||^2$): 
\begin{equation}
	dB = 2 {\rm Re}  p \ .
\end{equation}


\subsubsection{General solution} 
\label{ssub:gen}

In supergravity applications, it is usually straightforward to compute $dj$ and $d \omega$ directly from the spinorial equations imposed by supersymmetry. In this case, it is more convenient to compute first the torsions $p$ and $q$ in (\ref{eq:pq}) from the conformal Killing spinor equation (\ref{eq:cksa}). The computation involves the action of $\gamma_{mn}$:
\begin{equation}
	\begin{split}
		\gamma_{mn} \epsilon_+ = i j_{mn} \epsilon_+ - \omega_{mn} \epsilon_+^C \ , \\
		\gamma_{mn} \epsilon_+^C = -i j_{mn} \epsilon_+^C + \bar \omega_{mn} \epsilon_+ \ .
	\end{split}
\end{equation}
This allows us to rewrite (\ref{eq:cksa}) as\footnote{Writing (\ref{eq:cksa}) as $E^1_m \epsilon_+ + E^2_m \epsilon_+^C=0$, one would expect two vector equations; decomposing each into $(1,0)$ and $(0,1)$ parts would give four equations. However, $E^1_{0,1}$ and $E^2_{1,0}$ can be shown to be equivalent. This can also be seen from the fact that multiplying (\ref{eq:cksa}) by $\gamma^m$ is automatically zero, and from (\ref{eq:geeC}).}
\begin{equation}\label{eq:pqcks}
	p^A_{1,0}= 0 \ ,\qquad
	2 p^A_{0,1} + q_{1,0} \llcorner \bar \omega = 0 
	\ ,\qquad
	q_{0,1} = 0 \ .
\end{equation}
where $p^A_m \equiv p_m -i A_m$.

(\ref{eq:pqcks}) can also be translated into equations for the intrinsic torsions $w_i$ defined in (\ref{eq:wi}): 
\begin{subequations}\label{eq:ckswi}
	\begin{align}
		\label{eq:w3}w^3&=0 \ ,\\
			\label{eq:A10}
			i A_{1,0}&=  -\frac12 \overline{w^5_{0,1}} + \frac 14 w^4_{1,0} + \frac 12 \del B \ ,\\
			\label{eq:A01}
			i A_{0,1} &=  +\frac12 w^5_{0,1} -\frac34 w^4_{0,1} +\frac12 \bar\del B \ . 
	\end{align}
\end{subequations}
We see that (\ref{eq:A10}) and (\ref{eq:A01}) simply determine $A$ and do not impose any constraints on the geometry. On the other hand, $w^3=0$ has a geometrical meaning: namely, 
\begin{equation}\label{eq:cs}
	(d \omega)_{1,2} = 0 \ .
\end{equation}
When $\epsilon$ has no zeros anywhere, this is just a way of saying that the manifold $M_4$ should be \emph{complex}. 

Let us briefly review why\footnote{This idea is usually attributed to Andreotti.}. From its definition as a bispinor in (\ref{eq:bisp}), we know that the two-form $\omega$ is decomposable, i.e.~it can locally be written as a wedge of two one-forms: 
\begin{equation}\label{eq:omegadec}
	\omega=e^1\wedge e^2\ .
\end{equation}
These one-forms $e^i$ can be taken as generators of the holomorphic tangent bundle $T^{1,0}$; this defines an almost complex structure $I_\omega$. Clearly, if $I_\omega$ is integrable, $d \omega $ is a $(2,1)$-form, and hence (\ref{eq:cs}) holds. To see that the converse is also true, observe that (\ref{eq:cs}) can only be true if $d$ of a $(1,0)$ form never contains a $(0,2)$ part; or, by conjugation, if 
\begin{equation}\label{eq:deb}
	(d e^{\bar i})_{(2,0)}=0\ ,
\end{equation}
where $i=1,2$ is a holomorphic index.	
Consider now any two $(1,0)$ vectors $E_i$, $E_j$. We have the following chain of equalities:
\begin{equation}\label{eq:intproof}
([E_j, E_k]_{\rm Lie})\llcorner e^{\bar i} = [\{d, E_j\llcorner\}, E_k\llcorner] \llcorner e^{\bar i} = 
- E_k\llcorner \{ d , E_j\llcorner\} e^{\bar i}=-E_k \llcorner E_j\llcorner d e^{\bar i}=0 \ .
\end{equation}
In the first step, we have used Cartan's magic formulas relating $d$, Lie derivatives and vector contractions. (\ref{eq:intproof}) means that the Lie bracket of any two $(1,0)$ vectors is still $(1,0)$, which is the definition of integrability. So $I_\omega$ is a complex structure, and the manifold $M_4$ is complex. 

Conversely, if $M_4$ is complex, there exists a solution of (\ref{eq:cksa}). Given a complex structure $I$, let $\omega_I$ be a section of its canonical bundle $K\equiv \Lambda T^*_{1,0}$. $I$ defines a ${\rm Gl}(2,\cc)$ structure on $M_4$; but ${\rm Gl}(2,\cc)$ is homotopy equivalent to U(2), and for this reason there is actually a U(2) structure on $M_4$. This means that there always exists a two-form $j$ compatible with $\omega$, in the sense that $j\wedge \omega=0$ (as in (\ref{eq:SU2})), or in other words that $j$ is a $(1,1)$ for $I$; this also implies that $j$ and $I$ define together a metric via $g=I j$.\footnote{One can alternatively reason as follows. Given any metric $\tilde g$ on $M_4$, the `projected' metric $g_{mn}\equiv(\Pi_m{}^p \bar\Pi_n{}^q + \bar\Pi_m{}^p \Pi_n{}^q) \tilde g_{pq}= \frac12 (\tilde g_{mn}+ I_m{}^p I_n{}^q \tilde g_{pq})$ is hermitian with respect to $I$. One can then define a two-form via $J_{mn}\equiv I_m{}^p g_{pn}$, which is indeed antisymmetric, as one can easily check.} The volume form of this metric is just $-\frac12 j^2$; by choosing an appropriate function $B$, we can now define a normalized $\omega = e^{-B} \omega_I$ so that $\omega \wedge \bar\omega = 2 j^2$ is also true (again as in (\ref{eq:SU2})). We can now define the $w^i$ from (\ref{eq:wi}); since $I$ is complex, $w^3=0$. Finally, as remarked earlier, (\ref{eq:A10}) and (\ref{eq:A01}) simply determine $A$ in terms of the $w^i$ and $B$; it can be checked that it transforms as a connection. 

If $\epsilon$ has zeros $z_i$, only $M_4-\{ z_i\}$ will be complex, and not the whole of $M_4$. This is for example the case for $S^4$. As we discussed at the end of section \ref{sub:cksa0}, in this case a chiral conformal Killing spinor $\epsilon_+$ has a zero at one point; the complement of that point is conformally equivalent to $\rr^4$, which obviously admits a complex structure. Conversely, if one finds a complex structure on $M_4 - \{ z_i \}$, one can determine $A$ through (\ref{eq:A10}), (\ref{eq:A01}), and one should then check whether it extends smoothly to the entire $M_4$. 

\bigskip 

To summarize, the charged version of the conformal Killing spinor equation, (\ref{eq:cksa}), is much less restrictive than the uncharged version studied in section \ref{sub:cksa0}. We found that the only requirement on the geometry is (\ref{eq:cs}), which can be solved for example by requiring that the manifold is complex. Moreover, $A$ is determined in terms of the geometry by (\ref{eq:A10}), (\ref{eq:A01}).



\subsection{The $A\neq 0$ case in three dimensions} 
\label{sub:cksa3d}

In this section, we will deal with equation (\ref{eq:cksa}) in $d=3$. The arguments are very similar to those in $d=4$, and we will be brief. 

Given a spinor $\chi$, we can complete it to a basis with its complex conjugate:
\begin{equation}\label{eq:chibasis}
	\chi \ ,\qquad \chi^C \equiv C \chi^* \ ,
\end{equation}
where $C^{-1} \sigma_m C = - \sigma_m^T$. Any nowhere-vanishing $\chi$ defines an identity structure. We can indeed construct the bispinors
\begin{equation}\label{eq:bisp3d}
	\chi \otimes\chi^\dagger = \frac12 e^B \left ( e_3 - i  {\rm vol}_3 \right )
	\ ,\qquad
	\chi \otimes \overline{\chi} = - \frac i 2 e^B \, o\  \qquad 
	(o \equiv e_1 + i e_2)\ ,
\end{equation}
where $e_a$ are a vielbein for the metric on $M_3$. We defined $\overline{\chi} = \chi^T C^{-1}$. 
Notice  that in odd dimensions the map between bispinors and forms is not bijective; a bispinor can be identified both with an even or an odd differential form. In writing (\ref{eq:bisp3d}) we opted for odd forms.
In terms of this vielbein, one can also show
\begin{equation}\label{eq:sigmachi}
	\sigma^m \chi = e_3^m \chi - i o^m \chi^C \ ,\qquad m=1,2,3 \ . 
\end{equation}

We can now define ``intrinsic torsions'' by expanding $\nabla_m \chi$ in the basis (\ref{eq:chibasis}):
\begin{equation}\label{eq:pq3}
	\nabla_m \chi \equiv p_m \chi + q_m \chi^C\ .
\end{equation}
Alternatively, we can simply use the ``anholonomy coefficients'' $c^a{}_{bc}$ defined by $de^a = c^a{}_{bc} e^b \wedge e^c$. It is more convenient to work with $e^3$ and $o = e^1 + i e^2$, and to organize the $c^a{}_{bc}$ as
\begin{equation}
	\begin{split}
			d e_3  &\equiv {\rm Re} (w^1 e_3\wedge o) + i w^2\, o \wedge \bar o\ ,\\
			d o &\equiv w^3 e_3 \wedge \bar o + w^4 \, o \wedge \bar o 
			+ w^5 e_3 \wedge o \ . 
	\end{split}
\end{equation}
Here, $w^2$ is real, while all the other $w^i$ are complex, which gives a total of nine (which is the correct number for the $c^a{}_{bc}$). Together with $dB$, these are in one-to-one correspondence with the $p$ and $q$ in (\ref{eq:pq3}):
\begin{equation}\label{eq:pqw3}
	\begin{split}
		dB &= 2 {\rm Re}  p \ ,\qquad
		w^1 = - 2 i \bar q \cdot e_3 \ ,\qquad
		w^2 = {\rm Re} (q \cdot\bar o)\\
		w^3 &= i q\cdot o \ ,\qquad 
		w^4 = - i {\rm Im} p \cdot o \ ,\qquad
		w^5 =  i q \cdot \bar o + 2 i {\rm Im}  p \cdot e_3 \ . 
	\end{split}
\end{equation}
We are now ready to impose (\ref{eq:cksa}). Using (\ref{eq:pq3}) and (\ref{eq:sigmachi}), we get
\begin{equation}
	2 p^A \cdot e_3=  i q \cdot \bar o  \ ,\qquad
	p^A \cdot o = - 2 i q \cdot e_3 \ ,\qquad
	p^A \cdot \bar o = 0 = q \cdot o  \ .
\end{equation}
The first three simply determine $A$. The last can be written as $w^3=0$, which means that the sole geometrical constraint is that
\begin{equation}
	d o = w \wedge o
\end{equation}
for some $w$, in analogy to (\ref{eq:cs}). 



\section{Supersymmetric theories on curved spaces from new minimal supergravity} 
\label{sec:newmin}

In the previous section, we have studied the constraints imposed by the presence of at least one supercharge in a superconformal theory, by coupling the theory to conformal supergravity. We will now show that those results can be interpreted very naturally also as the coupling of a supersymmetric theory to ``new minimal supergravity'' \cite{sohnius-west, sohnius-west2}. In particular, we show that every solution of the new minimal equations is a conformal Killing spinor. Viceversa, every conformal Killing spinor (without zeros\footnote{We thank the authors of \cite{dumitrescu-festuccia-seiberg} for a  useful comment
about this point.}) gives rise to a 
solution of the new minimal equations. 
We can then use the results in section \ref{sec:cks} to understand when we can consistently define a  supersymmetric, but not necessarily conformal, theory  with an R-symmetry  on a curved manifold.   

\subsection{Equivalence with conformal Killing spinor equation} 
\label{sub:eq}

We start  with a solution of  the conformal Killing spinor equation (\ref{eq:cksa}) without zeros, charged under a connection $A$. 

As a first step, notice that $D^A \epsilon_+$ is a negative chirality spinor, and as such can be expanded in the basis (\ref{eq:-basis}): 
\begin{equation}\label{eq:vdef}
	D^A \epsilon_+ \equiv 2i v \epsilon_+
\end{equation}
where $v= v^m \gamma_m$ and $v^m$ is a vector.\footnote{Its $(1,0)$ part is immaterial because of (\ref{eq:proj}); its $(0,1)$ part can be written in terms of the intrinsic torsions in (\ref{eq:pq}) as $ v_{0,1}=-\frac i2\left( p^A_{0,1} - \frac12 q \llcorner \bar \omega\right)$.} Since  $\epsilon$ has no zeros, $v^m$ is  defined everywhere. 
 An easy computation now shows that (\ref{eq:cksa}) can be rewritten as\footnote{We use lower-case letters $a$ and $v$ for the auxiliary fields of new minimal supergravity, in order to avoid confusion with the $A$ of conformal supergravity we have been using until now.}
\begin{equation}\label{eq:nm}
	\nabla_m \epsilon_+ = -i \left( \frac12 v^n \gamma_{nm} + (v-a)_m\right) \epsilon_+ \ ,\qquad
	a \equiv A + \frac 32 v \ .
\end{equation}
This is exactly the condition for the existence of at least one unbroken supersymmetry in new minimal supergravity \cite{sohnius-west, sohnius-west2}. When this condition has a solution, we can consistently define supersymmetric theories on the four-manifold $M_4$ using the strategy in \cite{festuccia-seiberg}.

Actually, $v$ starts its life as the auxiliary field of a tensor multiplet; so one should impose that it can be dualized back: 
\begin{equation}\label{eq:d*v}
	d * v = 0 \ .
\end{equation} 
We can use the ambiguity in the definition of $v$ to arrange this condition. Since $v$ is defined only up to its $(1,0)$ part, we have two arbitrary complex parameters that can be used to enforce (\ref{eq:d*v}).
An alternative  geometrical perspective is the following. We can first choose $v$ imaginary and then  perform  a conformal rescaling of the metric
\begin{equation}
	g_{m n}\to e^{2f} g_{m n} \ \Rightarrow \ 
	D  \to  \left( e^{-f} D + \frac{d-1}2 \del_m f \gamma^m \right)\ .
\end{equation}
In $d=4$, this transforms $v \to v -i\frac34 d f $, and one can use the freedom in choosing $f$ to arrange so that (\ref{eq:d*v}) is satisfied. 

Hence we have shown that 
one can take the charged conformal Killing spinor equation (\ref{eq:cksa}) to the condition of unbroken supersymmetry in new minimal supergravity (\ref{eq:nm}). 
The fact that one can bring (\ref{eq:cksa}) to (\ref{eq:nm}) was to be expected because of the formalism of conformal compensators (for a review see \cite{vanproeyen-conformal}). In that formalism, one obtains new minimal supergravity by coupling a tensor multiplet to conformal supergravity, and by then giving an expectation value to the tensor multiplet. 

By reversing the previous argument,  it is clear that every solution of the new minimal equation (\ref{eq:nm}) is also a solution of the conformal Killing equation  (\ref{eq:cksa}).

We have now two ways of defining a conformal field theory on a curved background, either by coupling to  conformal supergravity or by coupling  to new minimal supergravity.  
The resulting theory is however the same. The coupling to new minimal supergravity will add linear and quadratic terms in the auxiliary fields $a,v$, as discussed in \cite{festuccia-seiberg}. 
At the linear level in the auxiliary fields the bosonic action contains the coupling to the supercurrent multiplets  \cite{sohnius-west,festuccia-seiberg}
\begin{equation}
-\frac 12 g_{mn}T^{mn} + \left (a_m -\frac 3 2 v_m \right ) J^m + \bar \psi_m {\cal J}^m  - \frac 1 2 b^{mn} t_{mn}\ ,
\end{equation}
where $J^m$ and $ {\cal J}^m $  are the R-symmetry and the supersymmetry current, respectively. In the non conformal case, the multiplet of currents also contains a conserved $t_{mn}$  ($\nabla^m t_{mn}=0$) which  measures precisely the failure of the theory at being conformally invariant. $t_{mn}$ couples to the dual of the auxiliary field $v$: $v^m=\epsilon^{mnpr}\partial_n b_{pr}$.   
In the conformal case $t_{mn}=0 $ and the linear coupling to $v_m$ vanishes. The remaining
terms  reproduce the couplings to the conformal supergravity (\ref{eq:cscoupling})  since $A= a-\frac32 v$.
The quadratic terms work similarly.


\subsection{One supercharge} 
\label{sub:reder}

Even though we have already analyzed the geometrical content of the conformal Killing spinor equation (\ref{eq:cksa}) in section \ref{sub:cksa}, it is instructive to repeat the analysis starting directly from (\ref{eq:nm}). 

We again introduce $j$ and $\omega$ as in (\ref{eq:bisp}). This time it is most convenient to calculate directly $dj$ and $d \omega$ from (\ref{eq:nm}). First of all we compute
\begin{equation}
	\label{eq:pure}
	\begin{split}
			d( \epsilon_+ \epsilon_+^\dagger)&=
			(-2{\rm Im} a \wedge + i {\rm Re} v\llcorner) \epsilon_+ \epsilon_+^\dagger + \frac12 e^B ({\rm Im}  v - i * {\rm Re}  v) \ ,\\ 
			d( \epsilon_+  \overline{\epsilon_+}) 
			&= 2 i a \wedge \epsilon_+ \overline{\epsilon_+}\ .
	\end{split}
\end{equation}
Using (\ref{eq:bisp}), we get expressions for $dj$ and $d \omega$, which in turn give us
\begin{subequations} \label{eq:nmeq}
	\begin{align}
		\label{eq:v01}
		v_{0,1}&= -\frac i2 w^4_{0,1} \ ,\\
		a_{0,1}&= -\frac i2 (\bar \del B + w^5_{0,1}) \ ,\\
		\label{eq:A10nm}
		\left(a - \frac 32 v\right)_{1,0} &= -\frac i4 w^4_{1,0} + 
		\frac i2 \overline{w^5_{0,1}} -\frac i2 \del B \ ,
	\end{align}
\end{subequations}
as well as 
\begin{equation}
	(d \omega)_{1,2}=0 \ . 
\end{equation}
Not surprisingly, these relations are consistent with (\ref{eq:ckswi}), which we found by directly analyzing the conformal Killing spinor equation (\ref{eq:cksa}). In particular, we have found again that the vector $A=\left(a-\frac32 v\right)$ is completely determined in terms of the geometry and $B$, and that the constraint on the geometry can be solved by taking the manifold complex, by following the steps described in section \ref{ssub:gen}. 

The vector $v$ must  satisfy $d *v = 0$. We can actually solve this condition for $v$ explicitly: although $A_{1,0}=\left(a-\frac32 v\right)_{1,0}$ is fixed by (\ref{eq:A10nm}), $a_{1,0}$ and $v_{1,0}$ are not. By choosing $a_{1,0}$ and $v_{1,0}$ in a convenient way we can  impose  (\ref{eq:d*v}). There is a particularly simple choice that always works. 
By choosing $a_{1,0}= -\frac i2 \del B + \frac i2 (w^4_{1,0}+ \overline{w^5_{0,1}})$, we get $v_{1,0}= \frac i2 w^4_{1,0}$, which together with (\ref{eq:v01}) gives
\begin{equation}\label{eq:vj}
	v= -\frac12 *dj\ .
\end{equation}
This obviously satisfies (\ref{eq:d*v}). 

Due to the ambiguity in choosing $a_{1,0}$ and $v_{1,0}$,  we can have different pairs $(a,v)$ that solve all constraints for supersymmetry.
For particular manifolds, for example $\mathbb{R}\times M_3$, there can be  different and more natural choices for $v$, as discussed below.

To summarize, using new minimal supergravity and the strategy in \cite{festuccia-seiberg}, a supersymmetric theory with an R-symmetry on any complex manifold $M_4$ preserves at least one supercharge. This is in agreement with our result in section \ref{sec:cks} for superconformal theories.

We will now comment in particular on the important subcase where $M_4$ is K\"ahler.

\subsubsection{ K\"ahler manifolds} 
\label{ssub:kahler}

A very simple case is  $v=0$ and $a$ real. The new minimal condition \ref{eq:nm} reduces to the equation for a covariantly constant charged spinor
\begin{equation}\label{eq:parallel}
 \left ( \nabla_m - i a_m \right ) \epsilon_+  =0 \ ,
\end{equation}
which is well known to characterize K\"ahler manifolds. In our formalism this can be seen easily from equations (\ref{eq:pure}). Using (\ref{eq:bisp}) we learn that $B$ is constant and
\begin{equation}
d j=0 \, , \qquad  d\omega = 2 i a\wedge \omega\ .
\end{equation}
The second condition implies, as already stressed, that $M_4$ is complex and the first that it is a K\"ahler manifold.  

It is interesting to consider the case   of conical K\"ahler  metrics
\begin{equation}
ds_4^2 = dr^2 + r^2 ds_{M_3}^2\ .
\end{equation}
The three dimensional manifold $M_3$ is, by definition, a Sasaki manifold.  The cone is conformally equivalent to the direct product  $\mathbb{R}\times M_3$ 
through the   Weyl rescaling $ds_4^2\rightarrow \frac 1 {r^2} ds_4^2$. 
$\mathbb{R}\times M_3$  will also support supersymmetry but with
different $a,v$. If we  keep the norm of the spinor fixed,  the new minimal conditions (\ref{eq:nmeq}) for a Weyl  rescaled metric
 $e^{2f} ds_4^2$ will be satisfied with the replacement
 \begin{equation}
 v\rightarrow v- i df\, \qquad  a\rightarrow a- i df\ .
 \end{equation}
In the  case of $\mathbb{R}\times M_3$,  we see that $a$ and $v$ have acquired an imaginary contribution  $i dt$ in terms of  the natural variable $r=e^t$ parameterizing $\mathbb{R}$. 
Notice that $v$ is  not of the form (\ref{eq:vj}) but that it nevertheless satisfies $d *v = 0$.

The theory on $\mathbb{R}\times M_3$  can be reduced to give a three dimensional supersymmetric theory on $M_3$.
 We will return to the study of  three dimensional theories in section \ref{sec:3d}.


  
\subsection{Two supercharges} 

It is interesting to consider the case where we have two supercharges $\epsilon_\pm$ of opposite chirality. The Euclidean spinors should
satisfy the new minimal equations \cite{festuccia-seiberg} which in the Euclidean read
\begin{equation}
	\begin{split}
		\label{eq:nm2}
			\nabla_m \epsilon_+ &= -i \left( \frac12 v^n \gamma_{nm} + (v-a)_m\right) \epsilon_+ \ , \\
			\nabla_m \epsilon_- &=  +i \left( \frac12 v^n \gamma_{nm} + (v-a)_m\right) \epsilon_- \ .
	\end{split}
\end{equation}

With two spinors, in addition to (\ref{eq:bisp}), we can construct the odd bispinors
\begin{equation}\label{eq:bisp2}
	\epsilon_+ \otimes\epsilon_-^\dagger = \frac14 e^B \left ( z_2 + * z_2\right )
	\ ,\qquad
	\epsilon_+ \otimes \overline{\epsilon_-} = \frac14 e^B \left ( z_1 + * z_1\right )
	\ ,
\end{equation}
where $\{z^i\}$ is a holomorphic vielbein, in terms of which $\omega= z^1\wedge z^2$ and $j= \frac i 2 \left ( z^1\wedge \overline{z^1} + z^2 \wedge \overline{z^2}\right )$.
It is easy to show that $z^1_m= \overline{\epsilon_-}\gamma_m \epsilon_+$ is a Killing vector. In fact
\begin{equation}
\nabla_{\{ m} z^1_{n \}} = 0\, , \qquad \nabla_{[ m} z^1_{n ]} = - i \epsilon_{m n}{}^{pr}  z^1_p v_r\ .
\end{equation}
We thus learn that we always have two isometries when there are two supercharges of opposite chirality. 

The commutator of the two supersymmetries closes on the isometry generated by $z^1$. For example, if we take the transformation rules for a chiral multiplet \cite{festuccia-seiberg,samtleben-tsimpis}:
\begin{subequations}
	\begin{align}
	\delta \phi &= - \overline{\epsilon_+}\psi_+\ ,  &   \delta \bar \phi &= - \overline{\epsilon_-}\psi_-\ ;  \\
	\delta \psi_+ &= F\epsilon_+  + \nabla^a_m \phi \gamma^m \epsilon_- \ , & \delta \psi_- &= \bar F\epsilon_+  - \nabla^a_m \bar \phi \gamma^m \epsilon_+\ ;\\
	\delta F &= \overline{\epsilon_-}\gamma^m \left ( \nabla_m^a -\frac i 2 v_m\right ) \psi_+ \ , &  \delta \bar F   &= \overline{\epsilon_+}\gamma^m \left ( \nabla_m^a +\frac i 2 v_m\right ) \psi_-\ ;
	\end{align}
\end{subequations}
a straightforward use of Fierz identities shows that
\begin{equation}
[ \delta_{\epsilon_+}, \delta_{\epsilon_-}] {\cal F} = {\cal L}_{z_1} {\cal F}\ ,
\end{equation}
where ${\cal F}$ is any field in the multiplet $(\phi,\psi_{\pm},F)$ and the Lie derivative ${\cal L}$ is covariantized with respect to $a$.



\section{New Minimal Supergravity reduced to Three Dimensions} 
\label{sec:3d}

It is also of some interest to reduce the condition of supersymmetry to three dimensions, where partition functions of Euclidean supersymmetric theories have been recently studied and computed using localization. 

In this section we will study the solutions of the  dimensionally reduced new minimal condition
\begin{equation}\label{eq:nm3d}
	\nabla_m  \chi = -i \left(  v^n \sigma_{nm} + (v-a)_m\right) \chi  +\frac{v_4}{2} \sigma_m \chi  \ ,\qquad
	m,n =1,2,3 \ ,
\end{equation}
where $\chi$ is a two-component three-dimensional spinor on the manifold $M_3$, $v$ and $a$ are vectors and $v_4$ is a scalar. Similarly to four dimensions, $v$ is subject to  the constraint $d(*v)=0$. A discussion of the off-shell ${\cal N}=2$ new minimal supergravity in three dimensions can be found  in \cite{kuzenko-lindstrom-tartaglino,kuzenko-tartaglino}.

Every solution of  (\ref{eq:nm3d}) can be uplifted to a solution  of the four-dimensional new minimal condition (\ref{eq:nm}) on a manifold with metric 
\begin{equation}\label{3d4d}
ds^2 =  e^{-2 b} ds_{M_3}^2 + e^{2 b} \left ( d\phi +\mu\right )^2 \, , 
\end{equation}
with connection $\mu$ determined by 
\begin{equation}
v+i \,db = \frac i 4 e^b * d\mu 
\end{equation}
and background fields $v_{4d}\equiv (v_4,v)$, $a_{4d}\equiv (v_4,a)$ satisfying $d(*v_{4d})=0$. We split all four-dimensional vectors in a component along $e^4= e^b(d\phi +\mu )$ and a vector on $M_3$.
We used the basis (\ref{eq:4dgamma}) and wrote the chiral spinor $\epsilon_+$  as $\epsilon_+ =
\bigl( \begin{smallmatrix}
			 \chi \\ 0 
\end{smallmatrix}\bigr)$. Notice that we identified $a_4=v_4$. 

More general reductions from four to three may exist and a more general analysis can be performed, but (\ref{eq:nm3d}) will be sufficient to illustrate various examples.

To characterize the geometry, we can use the bispinors we defined in (\ref{eq:bisp3d}). The equations they satisfy follow readily from the new minimal condition (\ref{eq:nm3d}):
\begin{align}
	\label{eq:3deq}
	d e_3 &= - (dB+ 2 \,{\rm Im} a) \wedge e_3 + 4 * {\rm Re} v + i\, {\rm Im} v_4\,  o \wedge \bar o \ , \\
	\label{eq:3deq2}
	d o &= (2\, v_4 e_3 + 2i\, a - dB)\wedge o\ , \\
	\label{eq:3deq3}
	dB &= 2\, {\rm Im} (v-a) + i \, {\rm Re} v  \llcorner ( o\wedge \bar o) + {\rm Re} v_4 e_3 \ .
\end{align}

As in four dimensions, the problem of finding solutions of the new minimal condition (\ref{eq:nm3d}) is closely related to the problem of finding solutions of the conformal Killing equation  (\ref{eq:cksa}).
In fact,  any  solution of  equation  (\ref{eq:cksa}) without zeros in $3d$ is also solution of   (\ref{eq:nm3d})  with the  scalar and vector $(v_4,v)$ defined by  $D^A \chi \equiv 3 \left ( i v^m\sigma_m +\frac 1 2 v_4\right ) \chi$
and $a= A + 2 v$. It should come as no surprise that all results obtained in section \ref{sub:cksa3d} are consistent and equivalent to the set of equations (\ref{eq:3deq}-\ref{eq:3deq3}). It is also obvious from this
discussion that not all components  of  the auxiliary fields $(v_4,v)$ are independent and there is some redundancy in their use.

We will now discuss some simple examples.

\subsection{Spheres, round and squashed}
\label{sub:S3}

Supersymmetric theories on the round sphere have been considered in \cite{kapustin-willett-yaakov,jafferis-Z,hama-hosomichi-lee}.
On $S^3$ we have Killing spinors satisfying $\nabla_m \epsilon= \pm  \frac 1 2 \sigma_m \epsilon$ and we can satisfy the new minimal condition
(\ref{eq:nm3d}) with $a=v=B=0$ and $v_4=\pm i$. It is easy to see how this translates in terms of forms.
We can define left- and right-invariant vielbeine:
\begin{equation}
ds^2 = \sum_{a=1}^3 l_a^2 = \sum_{a=1}^3 r_a^2\ .
\end{equation}
They satisfy $dl_a = \epsilon_{abc} l_b \wedge l_c$ and $dr_a = - \epsilon_{abc} r_b \wedge r_c$.  The equations (\ref{eq:3deq}) simplify to 
\begin{equation}  
d e_3 = 2 {\rm Im} v_4 \,e_1\wedge e_2\ \qquad d o = 2 v_4\, e_3 \wedge o\ ,
\end{equation}
which can be solved by taking $e_a$ to be a permutation of the $l_a$ for ${\rm Im} v_4=+1$ or a permutation of the $r_a$ for ${\rm Im} v_4=-1$.
It is worthwhile to notice that, if we define a superconformal theory on $S^3$, there will be no couplings linear in $v_4$ in the Lagrangian, and the theory will be invariant under all the supersymmetries with $\nabla_m \epsilon= \pm \frac12 \sigma_m \epsilon$; they obviously close to the superconformal algebra on $S^3$. If we instead consider a generic supersymmetric theory, $v_4$ will appear explicitly in the Lagrangian and we can only keep half of the supersymmetries.

One can also consider the squashed three-sphere
\begin{equation}
ds^2 = l_1^2 + l_2^2 +\frac{1}{s^2} l_3^2\ .
\end{equation}
Several different supersymmetric theories have been constructed on the squashed three-sphere \cite{hama-hosomichi-lee2,imamura-yokoyama} and have attracted some attention 
in the context of localization and the AGT correspondence \cite{alday-gaiotto-tachikawa}. For the interested reader, we quote the corresponding background fields. The simplest theory \cite{hama-hosomichi-lee2} is based on a deformation of the left invariant vielbein 
\begin{equation}
e_3=\frac {l_3} s \ ,\qquad e_1=l_1 \ ,\qquad e_2=l_2\ ,
\end{equation}
which corresponds to the
background fields $v_4=\frac is, a= \left(1-\frac 1 {s^2}\right ) l_3$ and $v=B=0$. A different theory has been constructed in \cite{imamura-yokoyama} and it is based instead
on a deformation of the right invariant vielbein
\begin{equation}
e_a = \cos \theta r_a + \sin \theta \epsilon_{abc} n_b r_c
\end{equation}
where $n_a$ is a unit vector on the sphere and $e^{i\theta} =\frac{1+i \sqrt{1-s^2}}{s}$.
The background fields are $v_4= -\frac i {2s}, a=v = \frac i 2 \frac {\sqrt{1-s^2}} s l_3 $ and $B=0$. 
The squashed three-sphere lifts to a four-dimensional bundle (\ref{3d4d}) with connection proportional to $v$, as also discussed in \cite{imamura-yokoyama}.  
The gravity dual of the theory in \cite{imamura-yokoyama} have been identified in \cite{martelli-sparks-nuts}, where also an analytical continuation of the theory, $(\theta,v,a)\rightarrow i  (\theta,v,a)$, has been considered.
One can explicitly check that the asymptotic behavior of the spinors in the gravity dual \cite{martelli-sparks-nuts} is consistent with our general discussion in section \ref{sub:43}.

Using the $v$ and $a$ computed for these examples to couple to the reduction of new minimal supergravity, one can check that one gets the same Lagrangians as in \cite{hama-hosomichi-lee2,imamura-yokoyama}.


\subsection{Sasaki Manifolds} 
\label{sub:squashed}

Another very general class of solutions is provided by Sasaki three-manifolds $M_S$. The spinorial characterization \cite{moroianu} of a Sasaki manifold is the existence of a solution of the
charged Killing equation
\begin{equation}\label{eq:sasaki1}
\left (\nabla_m - i a_m \right ) \chi = \frac{i}{2} \sigma_m \chi
\end{equation}
with real $a$. The new minimal condition (\ref{eq:nm3d}) provides a characterization in terms of a vielbein $\{ e_a \}$:
\begin{equation}\label{eq:Sforms}
d e_3 = 2 e_1\wedge e_2\, , \qquad\qquad d o = 2 i \alpha \wedge o \, .
\end{equation}
where $a= \alpha - e_3$. Notice that  $e_3$ has the property that $e_3 \wedge de_3$ is nowhere zero, which makes it a \emph{contact form} on $M_S$.

An equivalent characterization of a Sasaki manifold is the fact that the cone 
\begin{equation} ds_4^2 = dr^2 +r^2 ds_{M_S}^2
\end{equation}
is K\"ahler. Let us briefly review why the two characterizations are equivalent. First of all, it is easy to check that the spinorial equation (\ref{eq:sasaki1}) lifts  to the condition for a charged parallel spinor on the cone (equation (\ref{eq:parallel}))
whose existence is equivalent to the K\"ahler condition, as discussed in section \ref{ssub:kahler}. Alternatively, using the differential conditions (\ref{eq:Sforms}), we can construct a K\"ahler form $j =\frac 1 2 d (r^2 e_3)$ and a complex two form
$\omega = r (d r + i r e_3) \wedge o$   with $d\omega = w^5 \wedge \omega$.

We can generalize this example and include squashed Sasaki metrics \cite{herzlich-moroianu}
\begin{equation}
ds^2 = e_3^2 + \frac{1}{h^2} o_S\bar o_S
\end{equation}
where $e_3$ and $o_S$ satisfy (\ref{eq:Sforms}) and $h$ is a function with no component along the contact form ($e_3 \llcorner dh =0$). We can easily solve the new minimal conditions (\ref{eq:nm3d}) with background fields
$v_4= ih^2$ and
\begin{equation}
 a = \alpha - h^2 e_3 + \frac i {2h} \left ( \partial_{o_S} h - \partial_{\bar o_S} h\right )
\end{equation}
where  $ \partial_{o_S} h,  \partial_{\bar o_S} h$ are the components of $dh$ along $o_S$ and $\bar o_S$, respectively.
This class of manifolds is quite general and include for example Seifert manifolds, which are U(1) bundles over Riemann surfaces. On all these spaces we can easily define a supersymmetry field theory with at least a supercharge.

Notice that the new minimal condition here reads
\begin{equation}
\left (\nabla_m - i a_m \right ) \chi = i \frac{h^2}2 \sigma_m \chi\ .
\end{equation}
This kind of generalized Killing equations with a non-trivial functions on the right hand-side have solution only in dimensions less than or equal to three \cite{herzlich-moroianu}. 



\section*{Acknowledgments.}
This work is supported in part by INFN, the MIUR-FIRB grant RBFR10QS5J ``String Theory and Fundamental Interactions'', and by the MIUR-PRIN contract 2009-KHZKRX. We would like to thank D.~Cassani,  D.~Martelli  and L.~Girardello for useful discussions and the authors of \cite{dumitrescu-festuccia-seiberg}  for useful comments on our draft.



\providecommand{\href}[2]{#2}
\end{document}